# Asymmetric hysteresis for probing Dzyaloshinskii-Moriya interaction

Dong-Soo Han[1], Nam-Hui Kim[2,3], June-Seo Kim[1], Yuxiang Yin[1], Jung-Woo Koo[1], Jaehun Cho[2], Sukmock Lee[2], Mathias Kläui[3], Henk J. M. Swagten[1], Bert Koopmans[1], and Chun-Yeol You[2,4,*]

[1]Department of Applied Physics, Center for NanoMaterials, Eindhoven University of Technology, PO Box 513, 5600 MB Eindhoven, The Netherlands

[2]Department of Physics, Inha University, Incheon 22212, Republic of Korea

[3]Institut of Physics and Graduate School of Excellence Materials Science in Mainz, Johannes Gutenberg-Universität Mainz, 55099 Mainz, Germany

[4]Department of Emerging Materials Science, DGIST, Daegu, 42988, Republic of Korea

*Correspondence and requests for materials should be addressed to C.-Y.Y (email: cyyou@dgist.ac.kr)

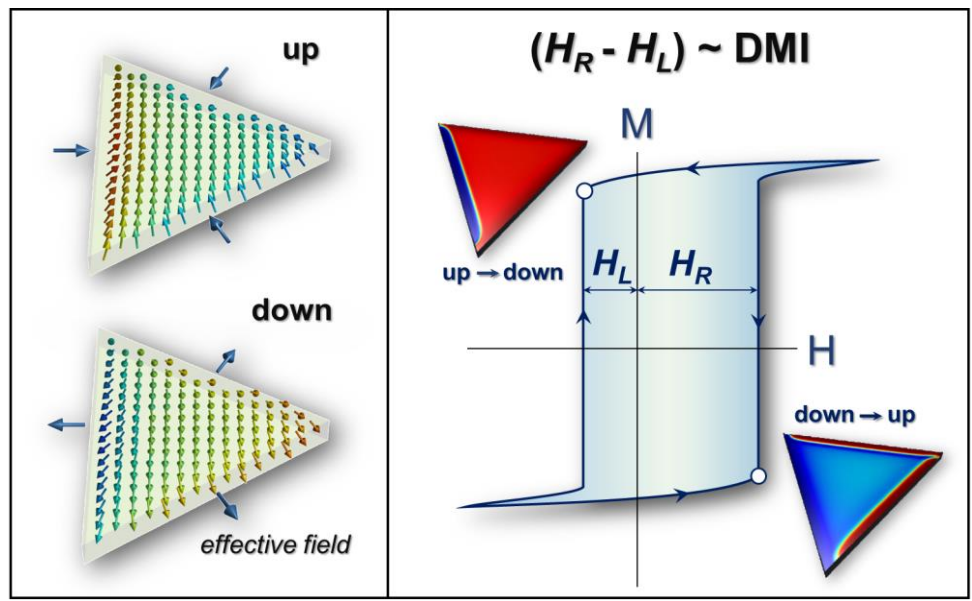

The interfacial Dzyaloshinskii-Moriya interaction (DMI) is intimately related to the prospect of superior domain-wall dynamics and the formation of magnetic skyrmions. Although some experimental efforts have been recently proposed to quantify these interactions and the underlying physics, it is still far from trivial to address the interfacial DMI. Inspired by the reported tilt of the magnetization of the side edge of a thin film structure, we here present a quasi-static, straightforward measurement tool. By using laterally asymmetric triangular-shaped microstructures, it is demonstrated that interfacial DMI combined with an in-plane magnetic field yields a unique and significant shift in magnetic hysteresis. By systematic variation of the shape of the triangular objects combined with a droplet model for domain nucleation, a robust value for the strength and sign of interfacial DMI is obtained. This method gives immediate and quantitative access to DMI, enabling a much faster exploration of new DMI systems for future nanotechnology.

Recently, the interfacial Dzyaloshinskii-Moriya interaction (DMI)[1-3] has been of great interest because of its intriguing physical role in the stabilization of nanometer-sized chiral spin configurations in perpendicularly magnetized multilayers, such as a Neel- type domain wall (DW)[4-10] and a nanomagnetic skyrmion[11-23], both very promising candidates for future information-processing devices with high-energy efficiency[6, 12, 14, 20] and ultrahigh-density storage capability[12, 13]. In particular, searching materials or systems providing strong DMI has become a central focus in current research, since it has been theoretically predicted[13] that room-temperature magnetic skyrmions may only be stabilized[15, 21, 22] by substantial DMI. To tackle the challenge met in the quest for large DMI systems, it is indispensable to develop a simple and accurate way of quantifying DMI. In this context, several experimental attempts have been made to estimate the sign and/or magnitude of the effective DMI coefficient $D$ through different measurement techniques. For out-of-plane magnetized systems, for instance, a number of ways to quantify the DMI have been suggested, mostly requiring imaging of magnetic domain walls[5, 24] and based on the analysis of domain wall propagation[6-10, 25-27] under the application of current or magnetic field. The asymmetric field-induced growth of reverse domains is the most widely used approach in this class of materials. However, this method has been found to entail serious drawbacks to quantify the DMI, as other effects, such as local stray field variations or anisotropy gradients, can also lead to an asymmetric displacement and analysis is hampered by possible additional effects resulting from wall spin structure deformations during the propagation. Furthermore, several asymmetric wall propagation observations[28-30] cannot be described with this model based on a pure DMI and it has been suggested that chiral damping[31], in addition to DMI, might play a key role. And even taking into account chiral damping, many of the observations[26] using domain wall displacement can still not be described, showing that a reliable quantitative determination of the phenomenon of DMI using *dynamic methods* is still tenuous.

Alternatively, measurements of *D* via non-reciprocal spin waves[32-40] have been recently realized in both in-plane and out-of-plane magnetized systems. Despite successful experimental demonstration of DMI by these approaches, it is still practically challenging to measure the magnetization dynamics in the ns or GHz regime, which requires unique measurement tools with high-spatial (wave vector) and/or time (frequency) resolution. Furthermore, the accuracy of this approach is also contested, as asymmetric anisotropies at the two interfaces of the ferromagnetic materials (FM) can lead to similar asymmetries of the dispersion relation,[41] making the interpretation for the DMI determination nontrivial. The technical complexities as well as issues with the reliability and robustness of these approaches are another hurdle to be overcome, impeding a rapid development of this exciting research field. Hence, in spite of the increasing interest to DMI-related phenomena, so far very limited experimental efforts have been devoted to extract DMI in films of (sub) nanometer thickness. And it is clear that to develop new materials, systems and devices, the main bottleneck is the reliable determination of the DMI. So there is an obvious urgent need to develop a novel *static method* that is not susceptible to chiral damping or asymmetric anisotropies, and that would allow for a straightforward and versatile determination of both the sign and magnitude of DMI.

In this work, we present the observation of a significantly shifted, asymmetric magnetic hysteresis loop manifested by interfacial DMI in laterally asymmetric microstructures. More specifically, in triangular-shaped ultrathin (~ nm) magnetic films patterns, where the lateral two-fold rotation symmetry with respect to the *z*-axis perpendicular to film plane is broken, we demonstrate a significant shift of the magnetic hysteresis along the magnetic field axis under the application of an additional static in-plane field, closely resembling the shifted 'exchange-biased' hysteresis[42, 43] due to coupling a ferromagnetic film with an antiferromagnetic material. This loop shift in our asymmetric triangular microstructures is found to reverse sign when changing the polarity of the in-plane field as well as when *D* changes sign, and moreover, it strongly depends on the strength of the in-plane field, all suggesting that a *unique straightforward* measurement of the DMI in nanoscale is feasible. Indeed, from a well-established half-droplet[29, 44, 45] model used to explain preferential nucleation at one of the sides of an unstructured nanomagnetic thin film, the shifted hysteresis loop is shown to directly yield the sign and strength of DMI. Moreover, the robustness and successful quantification of DMI is further substantiated by a systematic variation of the triangular shape of these structures, yielding shift in the hysteresis almost independent of the object size. This new tool is, thus, providing a fascinating prospect for a swift exploration of materials for nanoelectronics relying on engineered interfacial DMI.

In a confined system like a nano- or micro-structured magnetic thin film, DMI is known to allow for a boundary condition[13, 44, 46] of $2A\partial_{\mathbf{m}}/\partial_{\mathbf{n}} = D(\mathbf{z} \times \mathbf{n}) \times \mathbf{m}$, where **n** is unit vector normal to the side surface, *z* unit vector normal to film interface, **m** local magnetization orientation, *D* the Dzyaloshinskii-Moriya coefficient, and *A* exchange stiffness. Consequently, the magnetic configuration on the sideways facing surfaces (edges) of perpendicularly magnetized thin film structure experience an effective magnetic field, which tend to pull the local magnetization parallel or antiparallel to the surface normal vector **n** depending on the sign of $D$ and on the sign of the out-of-plane component of magnetization, $M_z$. This leads to an opposite tilted orientation of local magnetization on the opposite sides of the structure (see Figure 1a) over a length scale of the order of the domain wall width, which is typically a few nanometers. These chiral magnetic configurations at the sideways facing surfaces in a Pt/Co/AlO$_x$ magnetic thin film have recently been revealed by a pioneering experiment by Pizzini et al.[44], demonstrating chirality-induced asymmetric nucleation of magnetic domains when applying an in-plane bias field, $\mathbf{H}_x = \pm H_x \hat{x}$. According to their work, when the in-plane magnetic field is applied to the system, the chiral magnetic configurations at sideways facing surfaces give rise to a difference in energy barrier between the two opposite sides. Therefore, the nucleation process is found to be induced on one specific side where the Zeeman energy barrier is lowered when the tilt of the local magnetizations along the $x$-axis is parallel to $\mathbf{H}_x$.

A remarkable and useful aspect in the asymmetric switching mechanism is that the nucleation field is quantitatively related to the amount of Zeeman energy, which strongly depends on both the strength of $H_x$ and the orientation of local magnetization near the sideways facing surfaces relative to $\mathbf{H}_x$. In this regard, one would intuitively expect contrasting magnetic hysteresis loops within two different systems: one with laterally symmetric left and right sides and the other with antisymmetric sides, where the relative orientation of local magnetizations between two opposite sides are distinct. Figure 1a-b, for instance, schematically illustrate the magnetization distribution for $D > 0$ in two different structures with and without symmetric left and right side edges, respectively. In Figure 1c-d, their local magnetizations $\mathbf{m}_L$ ($\mathbf{m}_R$) at the left (right) side relative to each positive and negative $H_x$ are highlighted. In a square structure, when magnetization reversal occurs, the nucleation side dominating the switching process changes depending on the polarity of $M_z$ and $H_x$: for up-to-down (U-D) switching, the nucleation is initiated at the left (right) side for a positive (negative) $H_x$, vice versa for down-to-up (D-U) switching. However in this case, the local magnetization on one side with a specific $M_z$ shows an equivalent configuration as compared to the opposite side for an inverted $M_z$. As a result, the nucleation fields for both U-D and D-U switching should be exactly the same (see Figure 1c). This is in striking contrast to the case when switching a triangular object where the lateral two-fold rotation symmetry with respect to the *z*-axis perpendicular to film plane is broken. In this case, although the nucleation side also changes upon reversing the polarity of $M_z$, it experiences a different Zeeman energy at two opposite sides due to asymmetric orientation of local magnetizations at two opposite sides, yielding different nucleation fields for U-D and D-U. (see Figure 1d) Consequently, one can intuitively expect an essentially *asymmetric magnetic hysteresis loop* in the triangle, where it is shifted along the out-of-plane field axis under the application of the additional static in-plane field. We note that such an asymmetrically shifted hysteresis loop is a telltale sign of a symmetry-breaking interaction: usually magnetism is time-reversal invariant meaning that hysteresis loops are always symmetric around zero field (positive and negative fields have symmetric effects and are energetically degenerate). And by applying an in-plane field, this degeneracy is not lifted, thus still symmetric loops can be expected. Only effects, such as *exchange bias* can lift this degeneracy and are thus symmetry breaking[42, 43] and this effect relies on uncompensated spins at the interface between a ferromagnet and an antiferromagnet leading to an effective unidirectional bias field. However, we here show that the specific combinations of the device symmetry, the field symmetry and the symmetry breaking induced by the DMI result in a shifted hysteresis loop reminiscent of exchange bias and thus indicative of an effective bias field for the reversal, which is only slightly appreciated so far.

As a proof-of-concept and experimental demonstration of the feasibility of our approach, we carried out measurements on samples of Si(substrate)/Ta(4 nm)/Pt(4 nm)/Co(1.2 nm)/Ir(4 nm) and Si(substrate)/Ta(10 nm)/AlO$_x$(2.5 nm)/Co(1.15 nm)/Pt(4 nm) (hereafter denoted as Pt/Co/Ir and AlO$_x$/Co/Pt, respectively), with a strong perpendicular magnetization anisotropy (details, see Supplementary Information I and II). To implement symmetric and asymmetric left and right sideways facing edges, the magnetic thin films were micro fabricated into two patterns, *viz.* into a square and a triangle. (See Methods and Supplementary Information III). Magnetic hysteresis loops were monitored with a wide-field polar Kerr microscope applying a magnetic field of $\mu_0 H_z = -26.5$ mT into the out-of-plane direction to saturate the magnetization, and subsequently swept towards the opposite *z*-direction at a rate of 1 mT/s. During the measurement, an additional static in-plane bias field was subjected along the *x*-axis to initiate the chirality-induced asymmetric nucleation breaking the lateral two-fold symmetry. We would like to further note that square patterns, which should exhibit symmetric switching fields, were measured prior to triangle patterns, thereby excluding any possible artefacts which might arise from an unintentional misalignment of the in-plane magnet or exchange bias effects, which affect all geometries equally.

To validate the chirality-induced asymmetric nucleation process in our system – an essential prerequisite for the measurement of the DMI-shifted magnetic hysteresis loop – we first imaged magnetization dynamics close to the switching field. Figure 2 shows representative serial snapshots of instantaneous magnetic domain structures in the two different patterns as measured in a Pt/Co/Ir sample during its U-D and D-U switching (see Supplementary Information IV for AlO$_x$/Co/Pt). Here, the yellow-lined region indicates the micro-structured magnetic thin film, and the bright and dark contrast inside the structure represent negative and positive saturation of magnetizations, respectively. Two remarkable features are clearly seen in Figure 2. First, the nucleation side dominating the reversal process changes according to its switching polarity, clearly revealing chirality-induced asymmetric switching behavior as evidenced in previous work[44]. For D-U switching, the nucleation of magnetic domains is initiated at the right side while it starts at the left side for U-D switching for $\mu_0 H_x = +120$ mT in both the triangular and squared patterns. Accordingly, one can immediately deduce that our

Pt/Co/Ir sample stabilizes a left-handed chiral configuration with a moderate DMI, as in Pt/Co/AlO$_x$[15, 34, 36, 44]. Secondly, the nucleation field indeed differs depending on the side where the switching process is initiated in the triangle, whereas it is almost the same in the square. In the square pattern, the nucleation of magnetic domains appears at around $\mu_0 H_z \approx 2.9$ mT, irrespectively of the switching polarity. By contrast, it exhibits a significant and large difference (~1 mT) in the nucleation fields between the D-U and U-D switching in the triangular pattern. Here we note that statistically distributed small differences in the nucleation field between D-U and U-D switching were also observed in the squares due to the stochastic nature of domain nucleation as evidenced from repeatedly sweeping the magnetic field. However, these stochastic fluctuations are much smaller than the shifts in the triangles, where only systematic differences in nucleation fields were found upon repeatedly sweeping the field. In addition, the chiral switching was also confirmed upon reversing the polarity of $H_x$. In both patterns, completely opposite switching behaviour was found for a negative $H_x$, supporting the chiral and asymmetric switching behaviour in our samples (see the Supplementary Movies, demonstrating the dynamics of the nucleation and switching process).

To be able to quantitatively determine the interfacial DMI, we now discuss the magnetic hysteresis loops for these two distinctive geometries as is shown in Figure 3a-b. For Pt/Co/Ir, when no in-plane field is present, the square and triangular patterns show clear perpendicular magnetization with a coercivity $\mu_0 H_z \approx$ 17.5 mT and 18.4 mT, respectively. When the in-plane field is additionally introduced, *i.e.*, in this case $|\mu_0 H_x|$=100 mT, the square pattern exhibits a symmetric loop with reduced coercive fields of $|\mu_0 H_x| \approx$3.5mT, irrespectively of the polarity of $H_x$. In contrast, for the triangles indeed an asymmetric hysteresis loop with a positive coercive field $\mu_0 H_C^+ \approx$+7.7 mT and a negative coercive field $\mu_0 H_C^- \approx$-2.9 mT is observed for $\mu_0 H_x$= +100 mT. When reversing the in-plane field direction, *i.e.*, $\mu_0 H_x$= +100 mT, the loop is shifted into the opposite direction, perfectly in agreement with the concept of DMI-induced tilt of the local edge magnetization. Evidently, we successfully demonstrated asymmetric hysteresis loop with a significant horizontal offset of $|\mu_0 \Delta H_C| \equiv |H_C^+ + H_C^-| \approx 3.95$ mT corresponding to $\Delta H_C/H_0 \approx 21\%$ for $\mu_0 H_x = +100$ mT in the triangular-shaped Pt/Co/Ir sample.

For AlO$_x$/Co/Pt, similar features to the results from Pt/Co/Ir are seen (see again Figure 3). Interestingly we find for the reversed shifts of the magnetic hysteresis loops: at $\mu_0 H_x = +60$ mT, the magnetic hysteresis loop is biased to the left, but it is shifted to the right for $\mu_0 H_x = -60$ mT. Again, as the reduction or increase of Zeeman energy at the sideways facing surfaces is directly related to the sign the DMI, this contrasting response can be fully accounted for by the opposite tilt of local magnetization due to a different sign of $D$ between the two sample stacks. For AlO$_x$/Co/Pt, the platinum on top of the magnetic film is believed to play a dominant role in determining the sign of the DMI in this system, whereas Pt/Co/Ir includes it at the bottom of the magnetic film, suggesting an inverted contribution of Pt to the effective DMI[47] (see Supplementary Information IV for more detailed information).

In Figure 3 c-f, the variation of $H_C$ as a function of several in-plane magnetic fields $H_x$ is plotted. To be able to compare all system in the same scale and to extract the DMI (as discussed later on), the absolute values of $H_C$ for U-D and D-U are normalized to $H_0$, the coercivity without an in-plane field. The decrease of $|H_C/H_0|$ with increasing $|H_x|$ is seen over the entire field range due to the reduction in energy barrier for the magnetization reversal by the external field. In agreement with Figure 3a-b, for square patterns, symmetric features with respect to $H_x = 0$ are measured regardless of the switching polarity for both two samples. On the other hand, asymmetric curves are evidenced again in the triangles. For the case of Pt/Co/Ir, rapid (slow) decrease in $H_C/H_0$ with increasing $|H_x|$ for negative (positive) $H_x$ is measured for D-U, and vice versa for U-D. By contrast, for AlO$_x$/Co/Pt, due to the suggested inverted sign of $D$, obviously an opposite trend is seen in Figure 3d. To support the key aspects of our experimental data, we have performed micromagnetic simulations[48] as shown in Figure 3e-f

(see Method and Supplementary Information V). As seen in the Figure 3e, for $D = 2$ mJ/m$^2$, the general features of micromagnetic simulations are in excellent agreement with the experimental data, except for the quantitative value in $H_C/H_0$, which may originate from the discrepancy in material parameters and finite temperature effects in the experimentally measured samples. One of the most remarkable features here is that for $D = 0$, even in the asymmetric triangular structure, $H_C$ versus $H_x$ shows completely symmetric behaviour with respect to the $H_x = 0$. Hence, we can conclude that the measured asymmetric hysteresis loops are purely originating from the presence of a DMI-induced tilt of the local magnetization, making these shifted loops a unique and straightforward fingerprint for the existence of the DMI. Moreover, we like to emphasize that the possibility of an oxidation-induced exchange bias effect to explain our data is excluded: it has been experimentally demonstrated that, at the interface between Co/AlOx, over-oxidized Co may lead to an antiferromagnetic CoO and exhibit asymmetric hysteresis loops[49]. Although the shift in hysteresis loop from the exchange bias could occur in our system, it should be irrespective of the in-plane field and the shape of the microstructure, i.e., this would show up as a shifted hysteresis for $H_x = 0$ and for all geometries including square patterns. However, the measured shift in hysteresis loops shows a very specific dependence on the in-plane field sign and strength as well as the asymmetry in the patterned shape, inconsistent with an exchange bias effect.

As the asymmetric magnetic hysteresis loops are attributed to the relative orientation between local magnetizations near the sideways facing surfaces and the in-plane field, the shift of the hysteresis should necessarily be strongly dependent on the geometric angle of the pattern, i.e. the angle $\gamma$ indicated in Figure 4a. To examine the dependence of $H_C$ on the geometric angle of the triangle, we measured $H_C$ versus $H_x$ for three different cases $\gamma = 22.5°$, $30.0°$, and $37.5°$. Here, $H_C$ was chosen as the average of two switching fields of U-D and D-U for oppositely oriented in-plane fields. In all our samples, we found that the amount of asymmetry between positive and negative in-plane fields increases when making the triangle sharper, *i.e.* for smaller $\gamma$. (Figure 4b-c). This can be qualitatively explained in terms of the projection of the local magnetizations near the side edge onto the in-plane field axis, which increases with an increasing angle as $\sin(\gamma)$. Not only do these result additionally confirm the proof-of-concept of our tool for probing DMI, it also allows for a more accurate estimate on the strength of DMI, which will be discussed next.

In order to quantitatively address the measured angle dependence and explore the correlation between experimental data and *D*, we discuss the results on the basis of a well-established half-droplet model[44, 45, 50], which describes the nucleation field of a magnetic domain at the side edge under the application of both in-plane and out-of-plane fields (see Methods and Supplementary Information VI). Taking into account the geometric angle of a triangle, the normalized coercive field $H_{C,L}$ and $H_{C,R}$ which stands for the left- and right-side dominated nucleation process, respectively, can be written as:

$$H_{C,L}/H_0 = \sigma^2(H_x)/\left[\sqrt{1-(H_x/H_K)^2}\right]$$
$$H_{C,R}/H_0 = \sigma^2(H_x\sin(\gamma))/\left[\sigma^2(0)\sqrt{1-(H_x/H_K)^2}\right]$$
(1)

where the domain wall energy under the certain in-plane field with DMI is given as $\sigma(H_x) = \sigma_0\left[\sqrt{1-(H_x/H_K)^2} - (H_x/H_K + 2D/\sigma_0)\arccos(H_x/H_K)\right]$.[44] (see Supplementary Information VI). Here, $H_K$ is the perpendicular uniaxial anisotropy field, and $\sigma_0$ domain wall energy in the absence of an in-plane field and DMI, which is simply given by $\sigma_0 = 4\sqrt{AK_{eff}}$ with $K_{eff}$ the anisotropy resulting from the combined interfacial and shape anisotropies. We note that this analytical model is assuming that the Bloch wall profile partially introduced by the in-planed field parallel to the domain wall, is negligible for moderate DMI strengths (as shown in the Supplementary Information VI). Then, $H_C/H_0$ can be simply expressed in terms of the strength and sign of $H_x$ and the projection of the local magnetizations on the in-plane field axis – governed by $\sin(\gamma)$ – consistent with our earlier qualitative discussion. The best-fit curves to the experimental data using the above equations are shown in Fig. 4c-d by solid curves. The material parameters for $M_S$ and $H_K$ are taken from superconducting quantum interference device-vibrating sample magnetometer (SQUID-VSM) data as performed in unstructured Pt/Co/Ir and AlO$_x$/Co/Pt thin films. The exchange constant is assumed to be $A = 10$ pJ/m corresponding to the value used in a similar structure.[22] The analytically calculated data are generally in good agreement with experimental results, implying the adequacy of the half-droplet model for the present analysis, although some deviation from the equation is allowed due to our simplified approximation made in Eq. (1). Moreover, the fact that we can simultaneously describe our data for all four structures with different sharpness (defined by γ) by a single unique value of *D*, strongly supports the reliability of the method. From such a simultaneous non-linear least-squares fits to the results for all angles, we can extract $D = 1.69 \pm 0.03$ mJ/m$^2$ and $-1.43 \pm 0.06$ mJ/m$^2$ in Pt/Co/Ir (thickness of Co $t_{Co} = 1.2$ nm) and AlO$_x$/Co/Pt ($t_{Co} = 1.15$ nm), respectively, which corresponds to interfacial DMI constant $D_S = D \cdot t_{Co} = 2.03 \pm 0.04$ pJ/m and $-1.62 \pm 0.07$ pJ/m, respectively. Here we note that the simultaneous fit for all angles can give a much more accurate and reliable estimation on $D_S$, as compared to fitting results for a single angle case (we found up to ~15% variations in the estimate of $D_S$ between different angles). We also note that the estimated error in the (simultaneous) fit value is found to be less than 5 %, but further systematic errors could arise from the uncertainty in *A*, which we did not separately address in the current work. With two different exchange stiffnesses, $A = 5$ and $15$ pJ/m, which is in the range covered by other reports,[10, 11] we yield an approximately 25% deviation in the values for *D* as compared to $A = 10$ pJ/m. Table 1 summarizes $D_S$ as measured for similar samples through different techniques. For the case of AlO$_x$/Co/Pt sample, the magnitude of $D_S$ shows generally good agreement with those from other techniques, even considering the systematic errors from the uncertainty in $A$. For the Pt/Co/Ir sample, rather diverse values for $D_S$ are found probably due to structural (interfacial) differences between samples used for each measurement.

In order to better access the reliability of the estimated $D$'s, we also independently performed BLS measurement with the same material systems of Ta(4)/Pt(4)/Co($t_{Co}$)/Ir(4) and Ta(10)/AlO$_x$(2.5)/Co($t_{Co}$)/Pt(4) (nominal thickness in nm), where $t_{Co}$ varied from 0-2nm in a wedge shape (see Supplementary Information VIII). From various thicknesses of Co, we obtained $D_S$ = 1.61±0.14 pJ/m, and -1.15±0.12 pJ/m in Pt/Co/Ir, and AlO$_x$/Co/Pt, respectively, which are quite comparable to our determined values, implying that our quantification on $D$ is reliable in the present analysis and for this system, where the BLS method has been investigated thoroughly including possible artefacts from asymmetric interface anisotropies. Note that $D_S$ estimated in Pt/Co/Ir from both measurement techniques is slightly lower than $D_S$ = 2.20±0.02 pJ/m in Pt/Co/AlO$_x$, as measured in our previous study. Interestingly, this is in contrast to the theoretical prediction[47] as well as experimental data[22, 27]. In these studies, it is suggested that the effective DMI of Pt/Co/Ir is enhanced with respect to Pt/Co/AlO$_x$, as the sign of $D$ at Ir/Co is opposite to that for Pt/Co. In other words, Co sandwiched between Pt and Ir with an opposite sign of $D$ at each interface is expected to yield an increase of the total effective $D$. Our conflicting result could arise from many possible origins. One could speculate the enhancement of $D$ is due to the interface between Co/oxide. It has been theoretically predicted[15] that a Co/oxide interface could also give a significant additional contribution to the effective DMI. From recent *ab-initio* calculations, the magnitude of $D$ at Ir/Co interfaces[47] has been predicted to be very small and comparable to the value as predicted at the Co/oxide interface[15]. The large $D$ in Pt/Co/AlO$_x$, then, can be understood when a larger additional contribution from Co/AlO$_x$ interfaces arises as compared to the Pt/Co interfaces. Another possibility is that the $D$ at the Ir/Co interface has the same sign as that at Pt/Co, in contrast to previous results.[22, 27] Indeed, recent experiments on Ir/Co/AlO$_x$ performed by BLS[38] have shown that Ir/Co/AlO$_x$ has the same sign of *D* as compared to Pt/Co/AlO$_x$. The origin of this effect has not been fully understood, and obviously, different sample stacks and growth conditions may strongly influence the interfacial quality which in turn should be reflected in changes of the DMI.

In the present study, although only triangular-shaped microstructures were used as a prototype to demonstrate asymmetric magnetic hysteresis, we would like to note that this effect is also obtainable by introducing other shapes with a lateral asymmetry, such as a pentagon or a half-circular object. More importantly, our micromagnetic simulations performed for various lateral sizes of the triangles reveal that the shift in hysteresis loops is insensitive to the size of microstructures (see Supplementary information V), at least in a range of structures with lateral dimensions around 1-2 µm. This shape and size independence emphasizes the strength of our DMI probe in the possibility of handling a variety of microstructures including larger patterns for fast and reliable access. However, one should note that, in the latter case, the nucleation should predominantly take place at sideways facing edges instead of at local sites within the microstructure[44] or at the sharp edge point of the triangular structure. This requires sufficiently clean materials with sharp enough edges to make a meaningful analysis of DMI, which is rather standard in most deposition and processing of ultrathin magnetic films. Nevertheless, when dealing with considerable edge roughness e.g. creating a different nucleation barrier at each side of the triangle, we experimentally observed a rather different effect on the coercivity, although still affected by the presence of DMI (see Supplementary information VII). Next, the asymmetric magnetic hysteresis loops were quasi-statically demonstrated by using a wide-field Kerr microscope. However, more standard laboratory tools for the measurement of magnetic hysteresis loops, such as SQUID, VSM, and polar MOKE equipped with two-dimensional magnets, are also applicable to measure DMI, provided that a large enough signal-to-noise ratio is achieved by introducing a sufficiently large two-dimensional array of asymmetric magnetic objects. Finally, we would like to emphasize again that the interfacial DMI originating from atomic interactions at (sub) nanometer scale can be measured in a macroscopic manner without the need for any imaging technique or measuring its dynamics, providing a more straightforward and faster exploration of the DMI.

In summary, we establish a novel robust, simple and reliable approach to quantify the DMI in thin film structures. We successfully demonstrate a shift in magnetic hysteresis loops arising from the DMI, by introducing a lateral asymmetry in microstructures and applying an additional in-plane bias field. In our measurement, the presence of DMI and its sign can be straightforwardly determined in a robust manner by measuring the asymmetric behavior in switching fields, clearly depending on the polarity of the in-plane field and magnetization. Furthermore, with help of a half-droplet model, we are able to estimate the strength of DMI in two different sample stacks of Pt/Co/Ir and AlO$_x$/Pt/Co, which corresponds to $D_S$ = 2.03±0.04 pJ/m and -1.62±0.05 pJ/m, respectively, which is quite comparable to one obtained from other reported techniques. Our results open a novel avenue to a rapid and reliable screening of materials and systems for DMI, and thus overcome the key obstacle to design devices with tailored DMI for ultimate topological stability and efficient manipulation as required for next-generation devices for nanoelectronics.

**Method Part**

**Sample fabrication and characterization of magnetic properties.** We prepared square- and triangle-shaped magnetic thin films on a silicon wafer coated with 100-nm thick SiO$_2$ The magnetic thin films were deposited using DC magnetron sputtering with layer structures of Si(substrate)/ Ta(4)/Pt(4)/Co(1.2)/Ir(4), and Si(substrate)/ Ta (10)/AlO$_x$ (2.5)/Co (1.15)/ Pt(4), where the number in parentheses denotes nominal thicknesses in nm. The AlO$_x$ layer was obtained from plasma oxidation of Al layer as deposited on Ta. The plasma oxidation process was carried out for 10-min in an *in-situ* chamber at a 0.1 mbar background pressure of oxygen and a power of 15 W. The AlO$_x$/Co/Pt sample was annealed at 250º C for 30 minutes to introduce perpendicular magnetization. To improve pattern quality, we employed a PMMA/HSQ bilayer for a mask, prepared by the high-resolved electron beam lithography and polymer reactive ion etching (RIE), and Ar$^+$ ion milling is followed afterwards. Magnetic properties were measured on unstructured continuous films with superconducting quantum interference device-vibrating sample magnetometry (SQUID-VSM). The saturation magnetizations of ~920 and 1490 emu/cm$^{-3}$ were measured for Pt/Co/Ir and AlO$_x$/Co/Pt, respectively. All the films considered here exhibited perpendicular magnetic anisotropy (PMA), with their remanent state fully magnetized in the direction perpendicular to the substrate. The perpendicular uniaxial anisotropic field $\mu_0 H_K$ were measured around 0.7 T and 0.45 T for Pt/Co/Ir and AlO$_x$/Co/Pt, respectively.

**Magnetic hysteresis loops.** Magnetic hysteresis loops were acquired using a wide-field polar Kerr microscopy equipped with two-dimensional magnets. For the acquisition of reliable data set, the measurement of magnetic hysteresis loop was repeated over 10 cycles for each pattern, and 2-5 different patterns but with the nominally same shape were used for the measurement, and average of the data was used in Figure 4. The coercive field of each hysteresis loop was automatically extracted from a home-made code fitting of the normalized MOKE signal to $2(1/(1+\exp((H_z-A)/B)))-1$, where A and B are the fitting parameters. To get rid of artefacts which might arise from an unintentional misalignment of the in-plane magnet, square patterns, which should exhibit the symmetric switching fields, were measured before the measurements on triangle patterns.

**Micromagnetic simulations.** Micromagnetic simulations were carried out by using OOMMF code incorporating the interfacial DMI term. We used squares with lateral size of $L=1$ μm and triangles with $L=1$ μm and $\gamma=30º$ for the calculation. We also employed a unit cell size of

4×4×1.2 nm³ and the typical materials parameters for Co: saturation magnetization $M_S$=1090 kA/m, exchange stiffness $A$=10 pJ/m, and perpendicular magnetocrystalline anisotropy $K_u$=1.25 MJ/m, and a damping constant $\alpha$=0.5 to obtain rapid convergence for quasi-static investigations.

**Analytical model for the extraction of D value.** For rigid quantitative analysis of the experimental data, we adopted a half-droplet model for the estimation of nucleation field (see Supplementary Information). In the presence of DMI and an in-plane bias field $H_x$, the nucleation field at the edge can be expressed by

$$H_C = \frac{\pi t (\sigma_{DW})^2}{4\mu_0 M_S p k_B T \sqrt{1-(H_{in}/H_K)^2}} \quad (2)$$

Here, $\sigma_{DW}$ is domain wall energy density under the in-plane field and DMI, and $t$ is thickness of magnetic thin film, $\mu_0$ a magnetic permeability, $M_S$ saturation magnetization, $H_K$ in-plane saturation field, and $k_B T$ are Boltzmann constant and temperature. The $p$ is defined by the switching time $t$ simply expressed by $t = t_0 e^p$. In the experiments, the sweeping time of the out-of-plane field is fixed as 100ms (~1mT/s), and the switching time is typically in the order of ~ns, therefore, $p$ can be approximated to ~18. For a moderate DMI strength, the domain wall energy density $\sigma_{DW}$ can be approximated into

$$\sigma_{DW} = \sigma_0 \left( \sqrt{1-(H_{in}\cos\varphi/H_K)^2} + \left(H_{in}\cos\varphi/H_K + \frac{2D}{\sigma}\right)\left(\arcsin(H_{in}\cos\varphi/H_K) - \frac{\pi}{2}\right) \right) \quad (3)$$

where φ is azimuthal angle of the in-plane field with respect to the axis normal to domain wall.

The $D$ is estimated from the least-squares fit through experimental data of all angles in Figure 4b (or 4b) to Eq. (3), where $\varphi = \pi/2 - \gamma$.

**Brillouin light scattering.** The Brillouin light scattering spectra were independently performed. A $p$-polarized single longitudinal mode laser with a power of 300 mW and a wavelength of 532 nm was used as a light source. In order to measure Damon-Eshbach mode, an external static magnetic field was further parallel to the film surface. The back-scattered light from the sample is focused and collected. The *s*-polarized light is passed through the interferometer and the photomultiplier tubes. All measurements are performed at room temperature.

## ASSOCIATED CONTENT

**Supporting Information:**

Detailed sample preparation, characterizations, micromagnetic simulations, analytical derivation of the angle-resolved droplet model, BLS measurement, movies for the chirality-induced asymmetric switching.

## AUTHOR INFORMATION

**Corresponding Authors**

*(C.Y.Y.) E-mail : cyyou@dgist.ac.kr

**Author Contribution**

C.Y.Y. conceived the main idea. C.Y.Y, J.S.K, and D.S.H. contributed to the design of the experiment. D.S.H performed the sample fabrication, measurements with the Kerr microscope and analysis on the experimental data with the help from Y. Y and K.J.K.. N.H.K., J. C., and J.S.K. conducted BLS measurement and relevant data analysis. C.Y.Y and D.S.H. performed micromangnetic simulations and developed an analytical model. D.S.H wrote the manuscript with review and input from the other authors. H.J.M.S., B.K., and C.Y.Y. supervised the study. All authors discussed the results and commented on the manuscript.

**Notes**

The authors declare no competing financial interests.

**Acknowledgements**

We acknowledge supports from the research program of the Foundation for Fundamental Research on Matter (FOM), which is part of the Netherlands Organisation for Scientific Research (NWO), the National Research Foundation of Korea (Grant Nos. 2015M3D1A1070465 and 2014K2A5A6064900), the German Research Foundation (DFG via the joint Korean-German project KL1811/14 and the SFB TRR 173 Spin+X ) and the EU (MultiRev ERC-PoC-2014 665672).

**Figures**

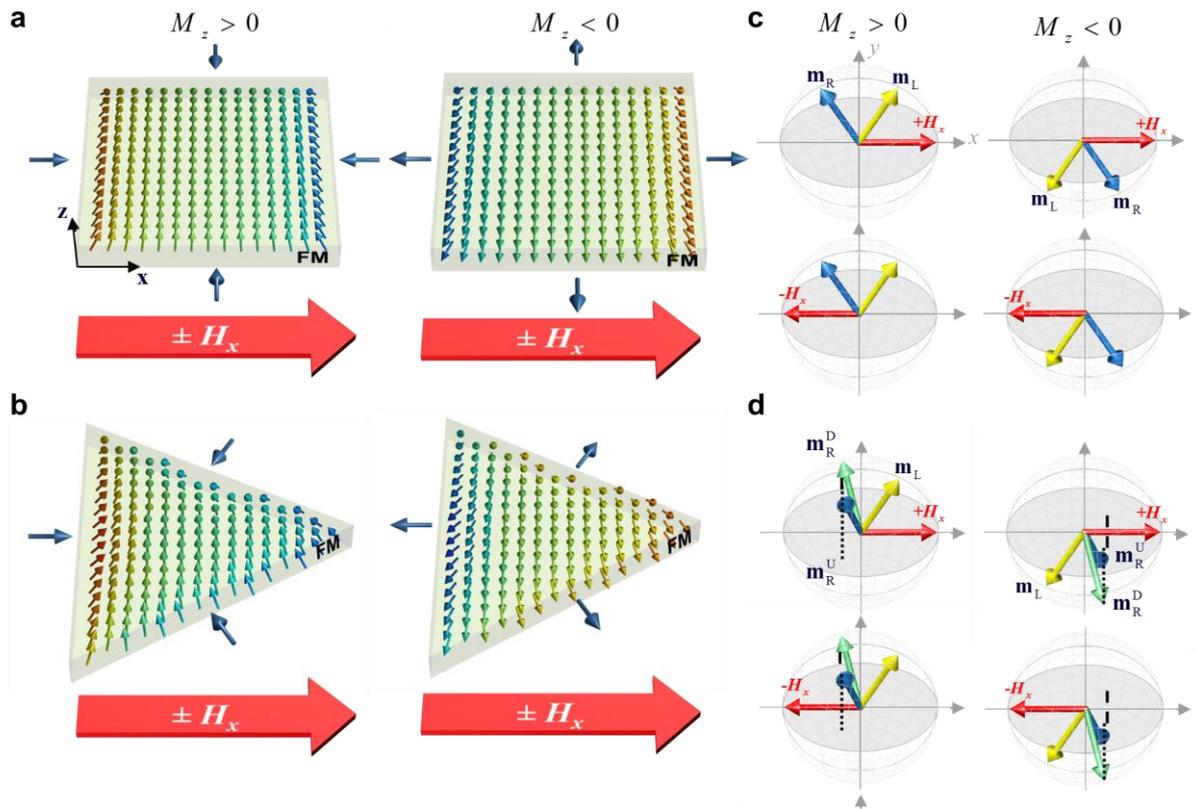

**Figure 1.** Schematic illustrations of magnetization configurations within a micro-structured ferromagnetic thin film with an interfacial DMI of $D>0$. The micro-structured ferromagnetic thin film consists of a square (a) and a triangle shape (b) which involve the lateral two-fold rotatory symmetry and asymmetry around $z$ axis, respectively. c and d represents the relative orientation between the local magnetizations at each left and right sideways facing surface and the in-plane field $\mathbf{H}_x$. The $\mathbf{m}_R$ and $\mathbf{m}_L$ indicates the local magnetization at the right and left side, respectively. The superscripts in each U and D represent the upper and lower side of the triangle structures, respectively.

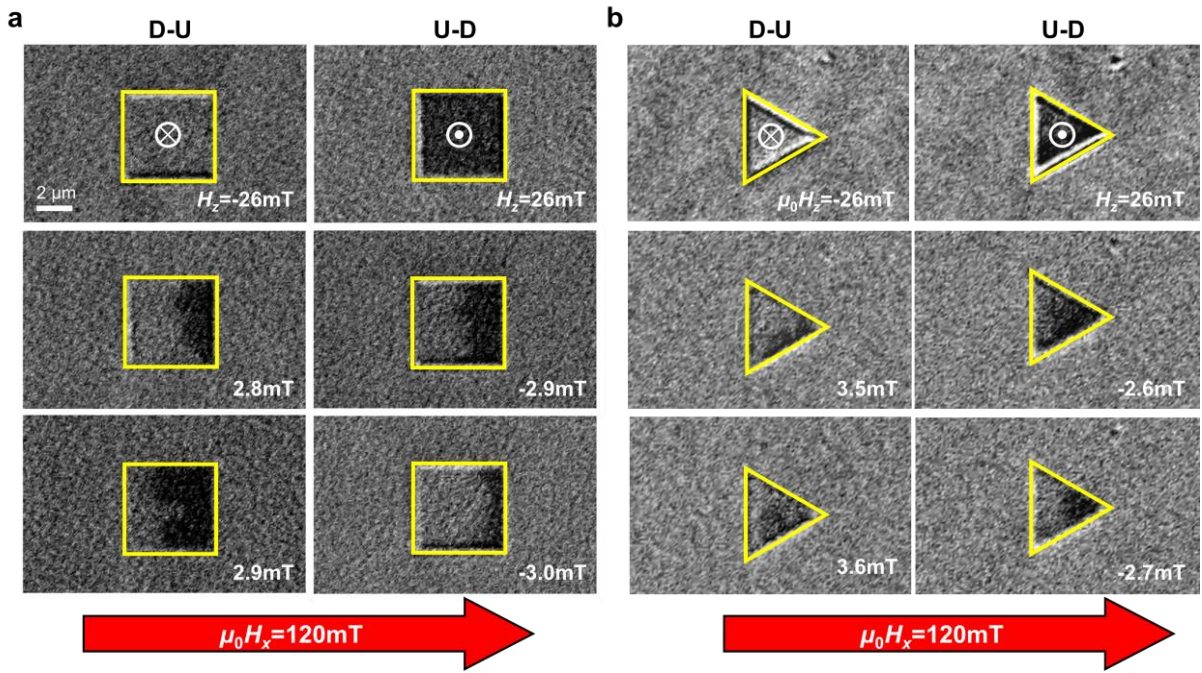

**Figure 2.** Serial snap shot images of local magnetic domain structure during its magnetization switching under the indicated in-plane $H_x$ and out-of-plane field $H_z$, as captured by the polar Kerr microscope in square (a) and triangle (b) patterns of Pt/Co/Ir. The left and right column in each panel of (a) and (b) represent the switching from down-to-up (D-U) and up-to-down (D-U), respectively. The yellow square indicates the micro-structured magnetic thin films. The grey and black colours in the structure represent negative and positive magnetizations along the z-axis.

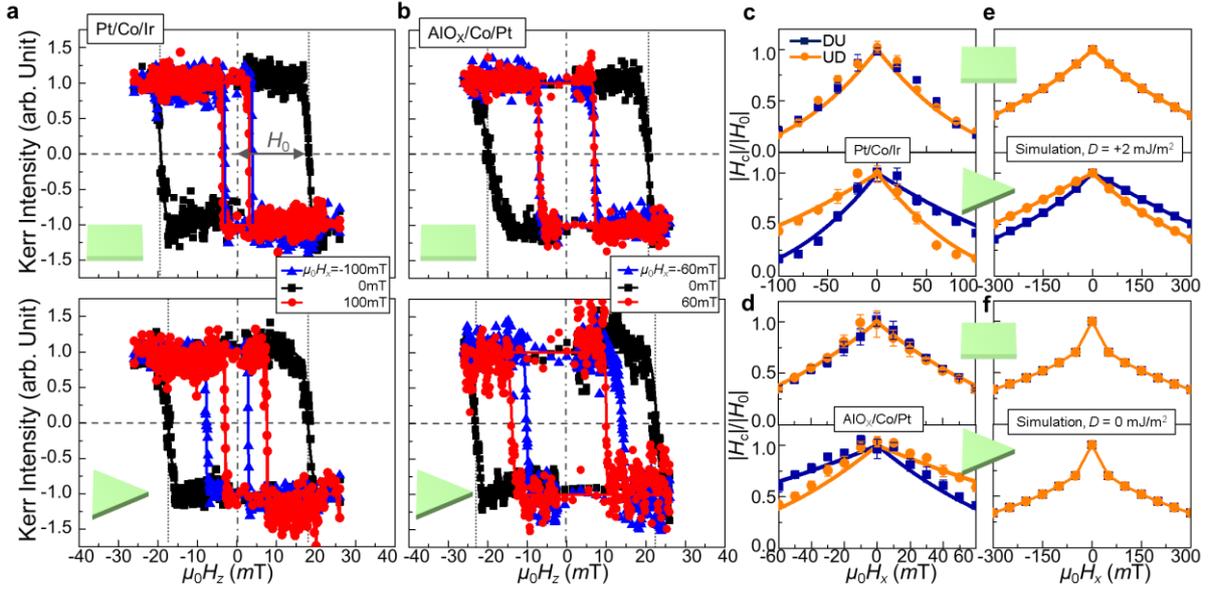

**Figure 3.** Magnetic hysteresis loops under the indicated in-plane bias field as measured in (a) Pt/Co/Ir and (b) AlO$_x$/Pt/Co. The symbols represent average experimental data measured in a single pattern over 10 times and solid lines indicates the least-squares fit to 2(1/(1+exp(($\mu_0H_z$-A)/B)))-1, where A and B are the fitting parameters to be determined. The top and bottom panels in each column of (a) and (b) represent the result from the square and triangle pattern, respectively. |H$_C$/H$_0$| versus $H_x$ as obtained from Pt/Co/Ir (c), AlO$_x$/Pt/Co (d), and micromagnetic simulations for $D$ = +2 (e) and 0 mJ/m$^2$ (f). The symbols in experimental cases represent data and solid lines are the least-square fit to Eq. 1. The top and bottom panel at each column of (c), (d), (e), and (f) are the results from square and triangle pattern, respectively.

.

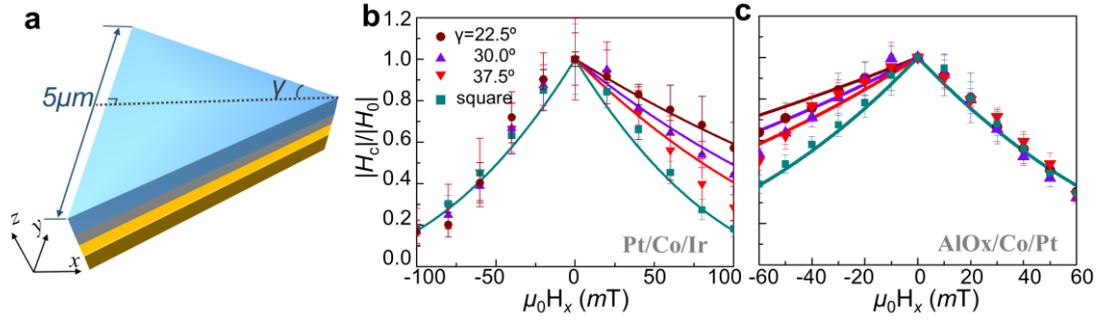

**Figure 4.** a. Schematic drawing of the triangle-shaped ferromagnetic thin film with indicated the vertical length of *5μm* and the angle *γ*. The triangle with the different angle *γ*= 22.5, 30, and 37.5⁰ were measured. Angular dependence of |$H_C/H_0$| versus $H_x$ in Pt/Co/Ir (b) and AlO$_x$/Co/Pt (c). The symbols represent average experimental data measured in 2-5 different patterns but with the nominally same shape and solid lines represent the least-square fit to Eq. 1.

### Table 1| Comparison of quatitative $D_s$ with other techniques

| Materials systems | $D_S$ (pJ/m) | Method | A (pJ/m) | Ref. |
|---|---|---|---|---|
| Pt(4)/Co(1-2)/AlO$_X$(2) | 1.43 | Brillouin Light Scattering | X | [34] |
| Pt(3)/Co(0.6)/AlO$_X$(2) | 1.32 | Chirality-iduced asymmetric nucleation | 16 | [44] |
| Ta(3)/Pt(3)/Co(0.6-1.2)/AlO$_X$(2)/Pt(3) | 1.72 | Brillouin Light Scattering | X | [36] |
| Ta(4)/Pt(4)/Co(1.4-2)/AlO$_X$(2)/Pt(3) | 2.20 | Brillouin Light Scattering | X | [37] |
| Ta(4)/AlOx(1.95)/Co(1-1.4)/Pt(4) | -1.15 | Brillouin Light Scattering | X | Supplementary Information |
| **Ta(4)/AlOx(1.95)/Co(1.15)/Pt(4)** | **-1.63** | **Asymmetric hysteresis** | 10 | |
| Pt(3)/Co(0.7)/Ir(0-1.3)/Pt(1) | -0.56 to 0.84 | Asymmetric magnetic domain growth | 16 | [27] |
| SiN/Pt(10)/Co(0.6)/Pt(1)/[Ir(1)/Co(0.6)/Pt(1)]$_{10}$/Pt(3)* | 1.15 | Imaging with a STXM | 10 | [22] |
| Ta(4)/Pt(4)/Co(0.8-1.6)/Ir(4) | 1.61 | Brillouin Light Scattering | X | Supplementary Information |
| **Ta(4)/Pt(4)/Co(1.2)/Ir(4)** | **2.05** | **Asymmetric hysteresis** | 10 | |
| Ta(4)/Ir(4)/Co(1.25-3)/AlO$_X$(2) | 0.84 | Brillouin Light Scattering | X | [38] |

The values in parentheses represent the nominal thickness of the layers in nanometers. The values of the exchange stiffness *A*, presented in the table, are based on the assumptions made in the literatures. 'X' in the column indicates that *A* is not used for extracting *D*.
*In the multilayer structure of [Ir/Co/Pt]$_{10}$, the thickness of a single Co layer is used to obtain $D_s=D*t_{FM}$.

# SUPPLEMENTARY INFROMATION

## I. OPTIMIZATION OF AlO$_X$/Co/Pt SAMPLE STACK

In order to find a proper growth condition for sufficiently strong perpendicular magnetic anisotropy (PMA) in AlO$_x$/Co/Pt, which is essential for the measurement of the asymmetric magnetic hysteresis loop, we investigated the coercive fields and remanence as a function of thickness of Co layer. The sample was grown on the 10 nm-thick Ta buffer layer to provide better interface quality, and the oxidation is carried out after the deposition of a 2.5nm-thick Al layer for 10 minutes in the *in-situ* oxidation chamber. The Co and Pt layers were deposited sequentially on the top of the oxidized Al layer by using DC magnetron sputtering. To find a optimal thickness for PMA, the Co layer is prepared in a wedge shape of which thickness ranges from 0 to 2 nm by using a linear shutter. The annealing process was additionally performed in an *ex-situ* furnace at 250º C for 30 minutes. Figure S1 shows coercive fields and remanence as measured from polar magneto-optical Kerr effect (MOKE) in the samples before and after the annealing process. As shown in Figure S1, for the case of as grown sample, no perpendicular magnetization is observed for whole Co thicknesses. By contrast, after annealing process, perpendicular magnetization starts to appear at *t*= 0.8 nm, and especially, strong PMA with 100% percent remanence from *t*= 0.9-1.3 nm is observed. In the present study, we chose *t*=1.15 nm which is close to the optimal thickness showing the coercive field of ~10 mT. After the patterning process, the coercive field is enhanced to ~18.5 mT probably due to the reduction of defects in the thin film, as seen in Figure 3 in the main text.

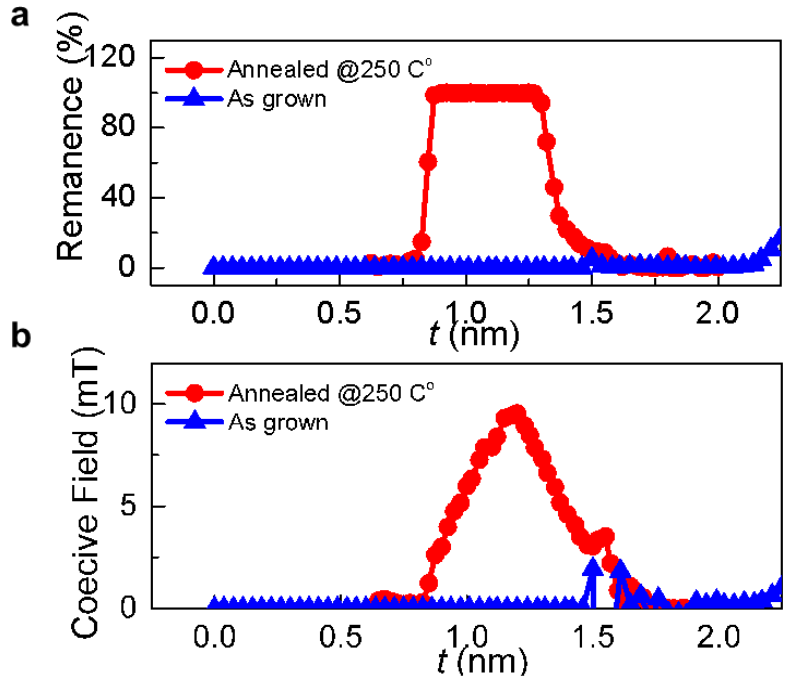

Figure S1. Remanence (a) and coercive field (b) as a function of cobalt thickness in a Ta(10 nm)/AlO$_X$(2.5 nm)/Co($t_{Co}$)/Pt(4 nm) wedge sample, measured by polar magneto-optical Kerr effect (MOKE) along the out-of-plane axis. The red and blue symbols denote the experiment data from the sample as-grown and after annealing at 250º C for 30 minutes, respectively.

## II. MAGNETIC CHARACTERIZATION OF THIN FILMS

To extract magnetic parameters of the thin films, we carried out magnetic characterization of thin films of Si(substrate)/Ta(4 nm)/Pt(4 nm)/Co(1.2 nm)/Ir(4 nm) and Si(substrate)/Ta(10 nm)/AlO$_X$(2.5 nm)/Co(1.15 nm)/Pt(4 nm), respectively, by using a SQUID-VSM magnetometer at room temperature. For AlO$_x$/Co/Pt, the sample annealed at 250º C for 30 minutes in an *ex-situ* furnace was used for the measurement, consistent with the sample as used in the main text. Figure S2 show the in-plane magnetic hysteresis loops of the samples. The magnetization $M_S$ and $\mu_0 H_K$ is extracted from the fitting experimental data to

$$2\left(1+\exp\left(\frac{\mu_0 H_z - A}{B}\right)\right)^{-1} - 1 \quad (S1)$$

where $A$ and $B$ are the fitting parameters to be determined. The magnetization $M_S$ approximately corresponding to 0.92 and 1.49 MA/m$^3$, and the in-plane saturation field $\mu_0 H_K$ of 0.7 and 0.45 T in Pt/Co/Ir and AlO$_x$/Co/Pt were measured, respectively. Here, a slightly low magnetization, compare to the bulk value of Co, ~1.4 MA/m$^3$, was observed in the Pt/Co/Ir, probably due to dead layers arising from intermixing between heavy metals and Co.

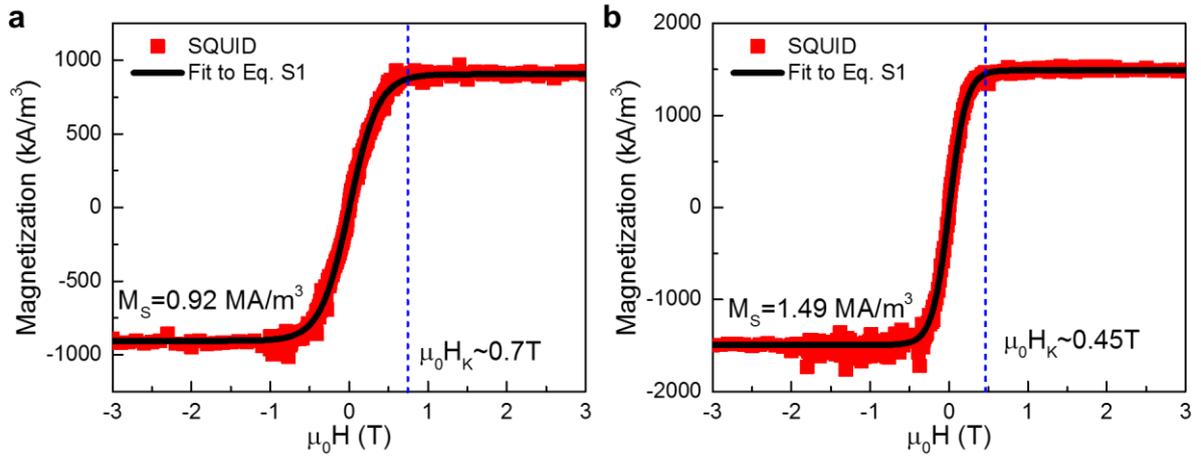

Figure S2. Magnetic hysteresis loops as measured by SQUID-VSM magnetometer along the in-plane axis in the **a** Pt/Co/Ir and **b** AlO$_x$/Co/Pt stacks. The black lines indicate fit to Eq. (S1)

## III. SAMPLE FABRICATION

In order to obtain well-defined microstructures of triangles and squares, we used highly resolved electron beam lithography and an $Ar^+$ ion beam milling technique. For a mask, PMMA/HSQ bilayers[1] which can provide both high resolution capability and good lift-off property are used. The overall process for the patterning is schematically illustrated in Figure S3a. The fabrication is started with spin-coting of PMMA 950K A2 on the sample and soft-baking at 170º C for 5 minutes on a hotplate. Afterward, HSQ is spin-coated on the top of the PMMA and soft-baked at 150º C for 2 minutes and 220º C for 2 minutes. Triangle and square patterns are drawn by an electron beam at 30 keV of an acceleration voltage. After the exposure to the electron beam, the sample is subsequently developed in tetramethyl ammonium hydroxide (TMAH) at room temperature for 90 seconds, rinsed in flowing DI water for 90 seconds and dried by blowing nitrogen gas. To imprint of the pattern in PMMA, the sample is further etched by the oxygen reactive ion etcher (RIE), and the $Ar^+$ ion beam milling process is followed afterward. In order to improve a signal-to-noise ratio in our optical signal, we removed the remained chemical mask by dissolving the sample in acetone. Finally, we obtain the patterns with very sharp edges, as seen in Figure S3b.

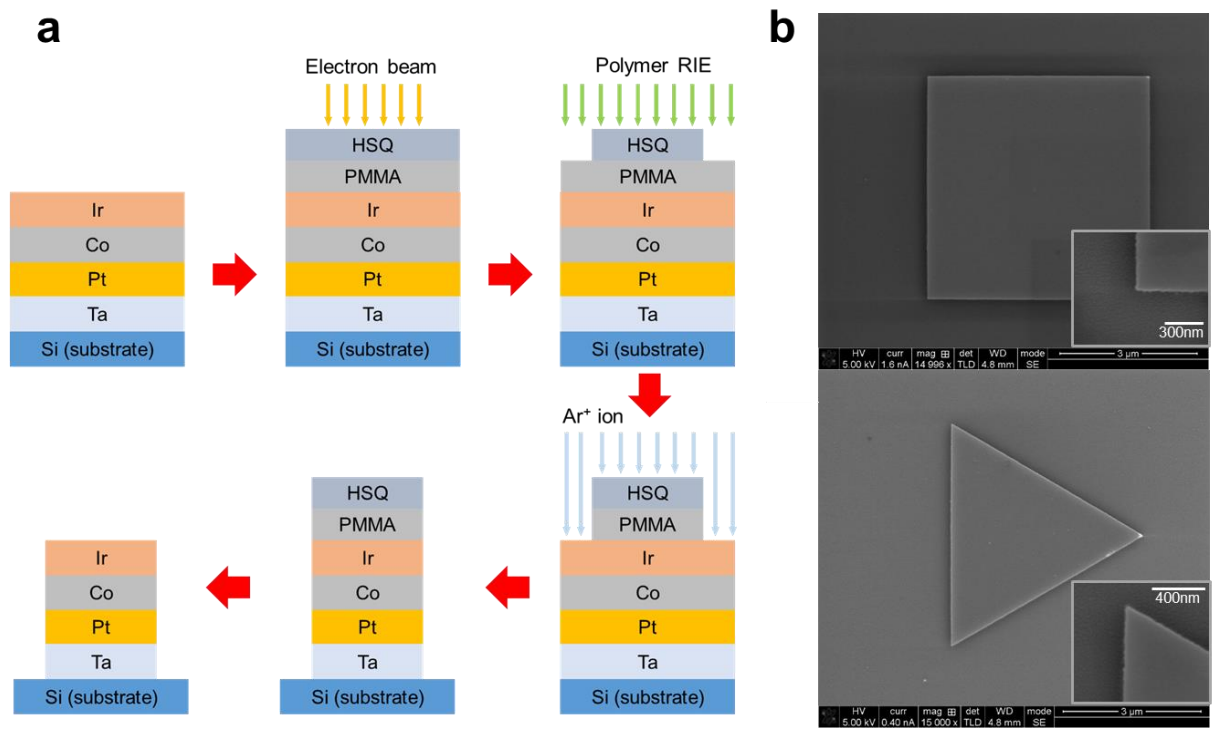

Figure S3. **a** Schematic illustration of patterning process by using HSQ/PMMA bilayer resists. **b** SEM images of square (top) and triangle (bottom) patterns from $Ar^+$ milling and the lift-off process.

## IV. CHIRALITY-INDUCED ASYMMETRIC SWITCHING IN AlO$_x$/Co/Pt SAMPLE

In this section, we discuss the chirality-induced asymmetric switching process, as measured in AlO$_x$/Co/Pt by a wide-field polar Kerr microscope. Figure S4 shows serial snapshot images of instantaneous magnetic domain structures in a square shape. The yellow-lined region indicates micro-structured magnetic thin film, and the bright and dark contrast inside of the structure represent negative and positive saturation of magnetizations, respectively. As seen in Figure S4, the asymmetric nucleation process depending on the switching polarity is clearly evidenced in a square. For the D-U switching, it is initiated at the left side, while it starts at the right side for U-D switching under $\mu_0 H_x$ = +80 mT. Such asymmetric nucleation behaviour for D-U and U-D switching is obviously opposite as compared to Pt/Co/Ir case in the main text, implying the different sign of *D* between two sample structures. We would like to note that, for the case of the triangular pattern, we could only measure a left-side-dominated nucleation process by the Kerr microscope due to a large difference in the nucleation field between two opposite sides. (see Supplementary Movies for more detailed information); for the case of the nucleation at the right side, the domain wall is subjected to the large $H_z$ because of its large nucleation field, the domain wall motion could not be measured by the present tool.

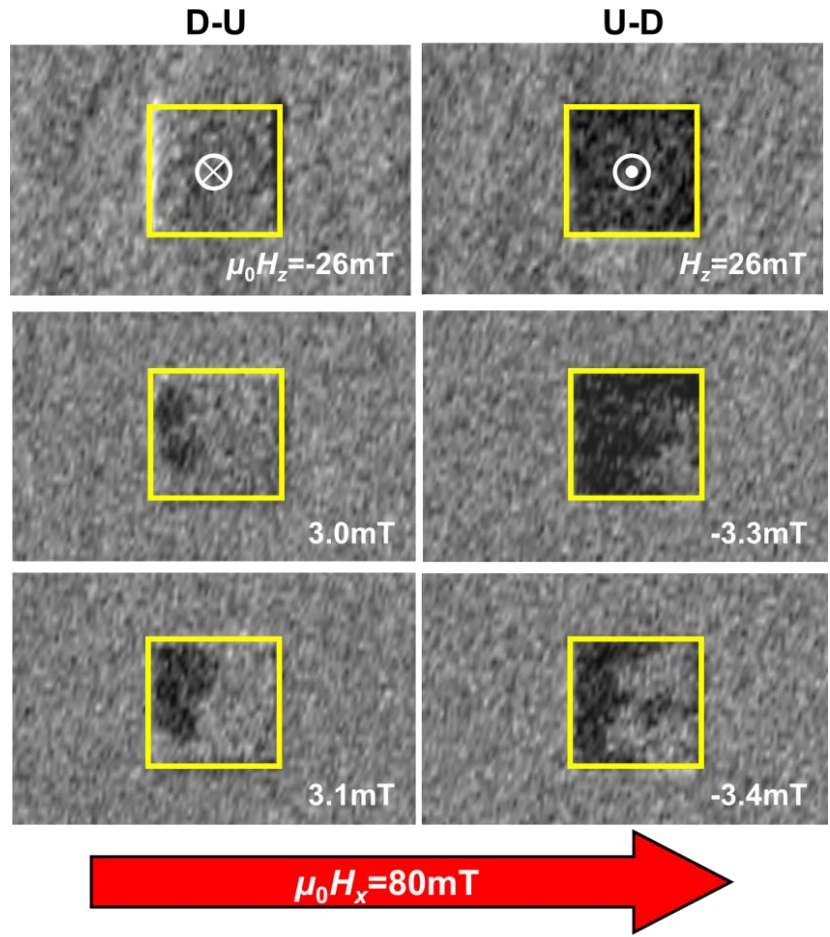

Figure S4 Serial snap shot images of local magnetic domain structure during its magnetization switching under the indicated in-plane $H_x$ and out-of-plane field $H_z$, as captured by the polar Kerr microscope in square patterns of $AlO_x/Co/Pt$. The left and right columns represent the switching from down-to-up (D-U) and up-to-down (U-D), respectively. The yellow-lined indicates the micro-structured magnetic thin films. The grey and black colours in the structure represent negative and positive magnetizations along the $z$-axis.

## V. MICROMAGNETIC SIMULATIONS

For qualitative analysis on our experimental data, here we performed micromagnetic simulations by using OOMMF code[2] which incorporates exchange, anisotropy, dipolar interaction, and DM interaction. In the calculation, in order to save time for calculations, we used triangles and squares with lateral sizes of $L$=1, 1.5, and 2 μm, which are smaller than the patterns as used in experiments. A unit cell size of $4\times4\times1.2$ nm$^3$, the typical materials parameters of saturation magnetization $M_S$=1090 kA/m, exchange stiffness $A$=10 pJ/m, PMA energy density $K_u$=1.25 MJ/m, effective DMI $D$=2.0 mJ/m$^2$, and damping constant α=0.5 were used in the simulations for fast convergence of our quasi-static studies. All simulations were calculated under zero temperature in which the thermal effect is neglected.

Figure S5 shows $H_C$ versus $H_x$ as obtained from three different geometric angle of $\gamma$= 15, 30, and 45º. The angular dependence of the asymmetry in hysteresis loop experimentally observed are successfully demonstrated in micromagnetic simulations, showing qualitatively similar behavior to experimental data. The difference in switching field between U-D and D-U reversal, $\Delta H_C = (H_C^+ + H_C^-)/2$, normalized to $H_0$, as displayed the inset of Figure S5a, clearly reveals the angle-dependent behavior; the asymmetry in hysteresis loop decreases with increasing geometric angle, and eventually it disappears at completely symmetric structure. Here, it is found that the qualitative value of the switching fields is quite different from the experimental data. This can be understood by thermal effect[3], which assists switching behavior, thereby leading to reduction in switching field. In Figure S5b, the asymmetry dependence on the size of the microstructure is further investigated. Interestingly, no significant difference in $\Delta H_C$ is seen in the results according to the change in their size, implying that the asymmetric magnetic hysteresis loop is insensitive to the size of microstructures.

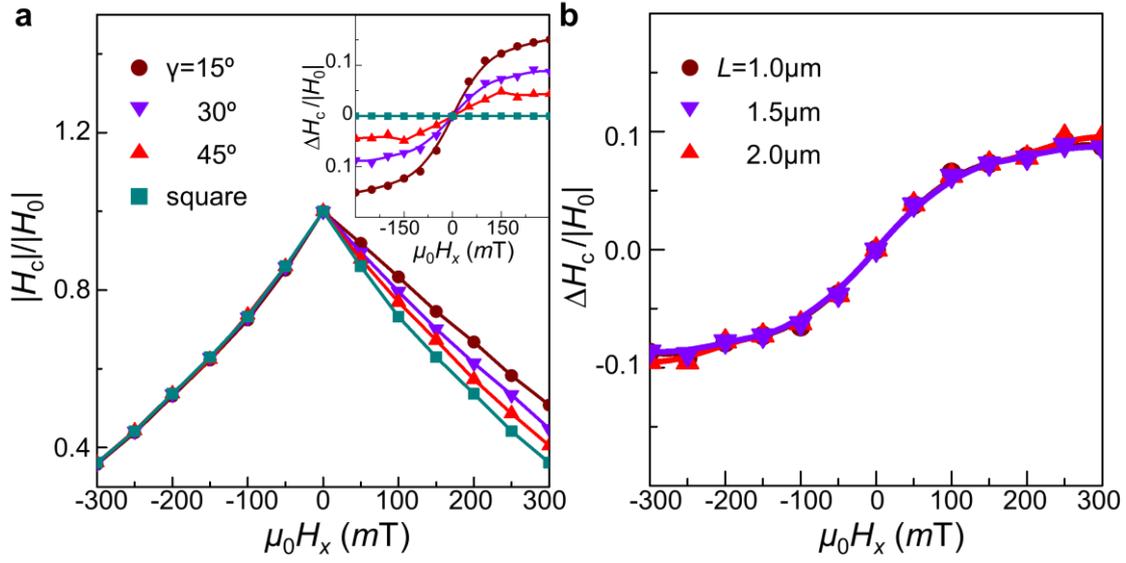

Figure S5. **a** Angular dependence of $|H_C/H_0|$ versus $H_x$ from micromagnetic simulation as conducted at $D=2$ mJ/m$^2$. The symbols represent micromagnetic simulation data and solid lines a guide to eyes. Inset in (a) shows $\Delta H_C/|H_0|$ versus $H_x$ for three different angles of triangle and square cases. **b** $\Delta H_C/|H_0|$ versus $H_x$ for $\gamma=30°$ and three lateral sizes $L$ of the triangle, where $L=1$, 1.5, and 2 μm.

## VI. ANALITCIAL EXPRESSION OF NUCLEATION FIELDS

A droplet model is a general way to describe a thermodynamic formation of a new phase or a structure through self-organization process. When a nucleation of a new phase arises by an external force, a small number of elements start to aggregate into a small cluster, here referred to as a droplet. In this case, a volume energy generally decreases with an increasing size of the droplet by the external force, while its surface energy increases. Accordingly, these two competing energies allow a local energy maximum at a certain size of the droplet, *viz* a *critical radius* ($r_c$), which corresponds to an energy barrier for the nucleation. For clusters with a smaller size than its critical radius, the total free energy of the system increases with increasing its size, while the total energy decreases when clusters are larger than the critical radius. Hence, the nucleation for the cluster with $r<r_c$ is prevented while that with $r>r_c$ is encouraged to expand its size.

The nucleation of a reversed magnetic domain in a magnetically saturated system can be also described in the framework of the thermodynamic droplet model by introducing Zeeman (magnetic volume energy) and domain wall energy (surface tension).[4] For the case of a uniformly magnetized domains, the volume energy only depends on the size of the droplet. By contrast, the internal magnetic configuration within a magnetic domain wall as well as the domain wall energy are markedly influenced by the interfacial DMI and the in-plane field, thereby, the energy barrier for the nucleation and the nucleation field exhibits strong dependence on the strength of the DMI and the in-plane field.

In the following sections, in order to estimate asymmetric coercive fields, we analytically discuss the dependence of the domain wall energy[3,5] and nucleation fields on the in-plane fields and DMI strength on the basis of the model used in Ref S3.

## 1. 1-Dimensional Domain Wall Model under an external in-plane field and DMI

For a one-dimensional regime, a Hamiltonian with Dzyaloshinskii-Moriya interaction (DMI) and Zeeman energy from an external in-plane magnetic field can be expressed by

$$E = \int_{-\infty}^{\infty} \left[ G_{ex}(\theta,\phi) - G_{DMI}(\theta,\phi) + \left[ G_{ani}(\theta,\phi) - G_{ani}(\theta_{\infty},\phi_{\infty}) \right] \right] dx \quad (S2)$$

where

$$G_{ex} = A \left[ \left( \frac{d\theta}{dx} \right)^2 + \sin^2\theta \left( \frac{d\phi}{dx} \right)^2 \right],$$

$$G_{DM} = D \left( \left( \frac{d\theta}{dx} \right) \cos\phi - \left( \frac{d\phi}{dx} \right) \sin\theta \cos\theta \sin\phi \right)$$

$$G(\theta,\phi) = K_{eff} \left( \sin^2\theta - \frac{2H_{in}}{H_K} \sin\theta \cos\phi \cos\varphi - \frac{2H_{in}}{H_K} \sin\theta \sin\phi \sin\varphi \right)$$

Here, $\theta$ and $\phi$ are polar and azimuthal angle of local magnetization in the wall, respectively, and $\varphi$ is azimuthal angle of the in-plane field with respect to the axis normal to domain wall. The first, second, and third term on the right side of Eq. (S2) is exchange, DMI, and anisotropy energy density, respectively. The anisotropy energy density includes perpendicular anisotropy and Zeeman energy. Generally, when the in-plane field applied with certain azimuthal angle, the variation in $\phi$ is introduced in the system, accordingly, Bloch wall profile is partially allowed in the wall. However, for a considerable DMI, since Neel-type domain wall is much favoured than Bloch wall by DMI and Zeeman energy by the in-plane field along the axis perpendicular to domain wall, the variation of $\phi$ in the wall is generally negligible. Then, we can assume that the domain wall energy from $\phi$ is negligible, accordingly, the Eq. (S2) can be approximated into

$$E = \int_{-\infty}^{\infty} \left[ A\left[\left(\frac{d\theta}{dx}\right)^2\right] - D\left(\frac{d\theta}{dx}\right) + \left[G_{ani}(\theta,\phi) - G_{ani}(\theta_\infty,\phi_\infty)\right] \right] dx \quad (S3)$$

where $G(\theta) = K_{eff}\left(\sin^2\theta - \frac{2H_{in}}{H_K}\sin\theta\cos\varphi\right)$.

From simple variational calculations, Eq. (S3) leads to

$$A\left(\frac{d^2\theta}{dx^2}\right) = K_{eff}\left(2\cos\theta\sin\theta - \frac{2H_{in}}{H_K}\cos\theta\cos\varphi\right) \quad (S4)$$

By multiplication with $d\theta/dx$ and integral of the equation with respect to $dx$, Eq. (S4) is given by

$$A\left(\frac{d\theta}{dx}\right)^2 = K_{eff}\left(\sin^2\theta - \frac{2H_{in}}{H_K}\sin\theta\cos\varphi\right) + C_1 \quad (S5)$$

For a magnetic domain wall in the infinite medium, the magnetization is stationary. Then, Eq. (S5) allows a hidden boundary condition that $d\theta/dx = 0$, accordingly, the integral constant $C_1$ can be expressed as

$$C_1 = -K_{eff}\left(\sin^2\theta_0 - \frac{2H_{in}}{H_K}\sin\theta_0\cos\varphi\right) \quad \text{where } \theta(x=\infty) = \theta_0 \quad (S6)$$

The stationary condition allows an extra condition that the total energy density should be stationary at infinite position. In this case, the first and second terms in Eq. (S3) naturally approach zero because $d\theta/dx = 0$, therefore, the stationary condition result in the condition that the anisotropy functional should be stationary with respect to $\theta$, leading to

$$\sin\theta_0 = \frac{H_{in}}{H_K}\cos\varphi \tag{S7}$$

The Eq. (S6) and (S5) indicate that the exchange energy density at certain $\theta$ in the wall is exactly the same with anisotropy energy density. Therefore, the domain wall energy density can be simply given by

$$\sigma_{DW} = \Delta\int_{\theta_0}^{\pi-\theta_0}\left[2K_{eff}(\sin\theta-\sin\theta_0)^2 \mp D\frac{(\sin\theta-\sin\theta_0)}{\Delta}\right]\frac{d\theta}{(\sin\theta-\sin\theta_0)} \tag{S8}$$

From the integration of Eq. (S8), we finally obtain

$$\sigma_{DW} = \sigma_0\left(\sqrt{1-(H_{in}\cos\varphi/H_K)^2} + \left(H_{in}\cos\varphi/H_K + \frac{2D}{\sigma}\right)(\arccos(H_{in}\cos\varphi/H_K))\right) \tag{S9}$$

Here, $\sigma_0$ is domain wall energy density under no in-plane field and DMI, which is simply given by $\sigma_0 = 4\sqrt{A\,K_{eff}}$.

To validate our approximation made in Eq. (S9), we further numerically calculated the domain wall energy in Eq. (S2) with consideration of both $\theta$ and $\phi$ variation in a magnetic domain wall, and compared the results with those from Eq. (S9). The calculations are performed by using typical material parameters of Co; $M_S$=1090 kA/m, exchange stiffness $A$=10 pJ/m, and perpendicular anisotropy field $H_K$=1 T, and effective DMI $D$=2.0 mJ/m². Figure S6a and S6b shows the domain wall energy $\sigma_{DW}$ normalized to $\sigma_0$ for various $\varphi$ and $H_{in}/H_K$, as obtained from numerical calculation of Eq.(S2) and Eq. (S9), respectively. The overall feature from Eq. (S2) is quite similar to the data from Eq. (S9), providing the reliability of our approximation in the Eq. (S9), although it is seen that the domain wall energies from Eq. (S2) is slightly higher than those from Eq. (S9) for very large angle $\varphi$. For a better comparison between them, in Figure S6c, $\sigma_{DW}/\sigma_0$ versus $H_{in}/H_K$ for selected angles of $\varphi$ =52.5, 60, and 67.5º, which corresponds to the geometric angle of $\gamma$ =37.5, 30, and 22.5º, is also compared.

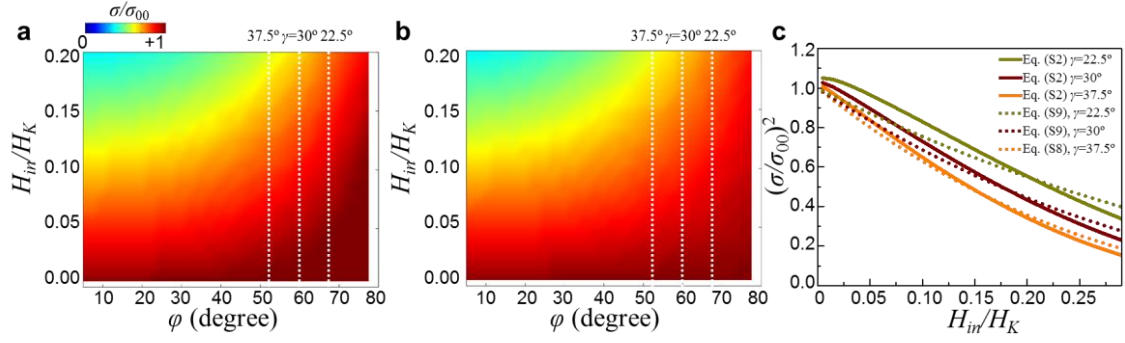

Figure S6. **a, b** Numerically calculated domain wall energies from Eq. (S2). and (S9) as a function of azimuthal angle of in-plane field $\varphi$ and $H_{in}/H_K$. **c** Comparison between **a** and **b** for selected angles of $\gamma = 37.5, 30,$ and $22.5°$.

## 2. A half-droplet model for estimation of nucleation fields

**Rigid droplet model**

Here, we discuss a droplet model[3-5] to estimate nucleation field at the edge of the microstructure. For the shake of simplicity, here, we adopted a rigid half-droplet model, which describes an isotropic and homogeneous nucleation process, as discussed in Ref. S3. For a rigid droplet model, the free energy of the droplet under the application of an in-plane and out-of-plane field is simply given by

$$\Delta E = \pi R t \sigma_{DW} - \pi R^2 t \mu_0 M_s \left( \sqrt{1 - \left(H_{in}/(H_K + H_z)\right)^2} H_z \right) \quad (S10)$$

Here, the first and second terms on the right side of Eq. (S10) indicate surface energy between two magnetic domains and volume energy, respectively. As described above, the energy barrier for the nucleation is determined by a total free energy at the critical radius where the free energy of the droplet corresponds to a maximum, and this corresponds to

$$E = \frac{\pi (\sigma_{DW})^2 t}{4 \mu_0 M_s \sqrt{1 - \left(H_{in}/(H_K + H_z)\right)^2} H_z} \quad (S11)$$

Here $\sigma_{DW}$ is the domain wall energy as derived in Eq. (S9). At a finite temperature, the probability of the nucleation and the nucleation rate can be simply expressed by an Arrhenius equation which is described by thermal activation over an energy barrier. In consideration of

the energy barrier and the thermal effect, a switching time $\tau$ is typically expressed as

$$\tau = \tau_0 \exp\left(\frac{-E}{k_B T}\right) \qquad (S12)$$

where $E$ is the energy barrier for the nucleation, $\tau_0$ initial switching time, $k_B$ Boltzmann coefficient, and $T$ temperature. Assuming that $p \equiv E/k_B T$ and the nucleation field is much lower than anisotropy field $H_K$, the nucleation field at side edges can be approximated into

$$H_C = \frac{\pi t (\sigma_{DW})^2}{4\mu_0 M_s p k_B T \sqrt{1-(H_{in}/H_K)^2}} \qquad (S13)$$

For the case of triangle structures, $\sigma_{DW}$ is different for left and right side because of its different tilting orientation, accordingly the nucleation fields $H_{C,L}$ and $H_{C,R}$ at the left and right side, respectively, can be given by

$$H_{C,L} = \frac{\pi t \left(\sigma_0 \left(\sqrt{1-(H_{in}/H_K)^2} + \left(H_{in}/H_K + \frac{2D}{\sigma}\right)\left(\arcsin(H_{in}/H_K) - \frac{\pi}{2}\right)\right)\right)^2}{4\mu_0 M_s p k_B T \sqrt{1-(H_{in}/H_K)^2}} \qquad (S14)$$

$$H_{C,R} = \frac{\pi t \left(\sigma_0 \left(\sqrt{1-(H_{in}\cos\varphi/H_K)^2} + \left(H_{in}\cos\varphi/H_K + \frac{2D}{\sigma}\right)\left(\arcsin(H_{in}\cos\varphi/H_K) - \frac{\pi}{2}\right)\right)\right)^2}{4\mu_0 M_s p k_B T \sqrt{1-(H_{in}/H_K)^2}}$$

(S15)

# VII. NON-IDENTICAL INITIAL POTENTIAL BARRIER EFFECT ON ASYMMETRIC HYSTERESIS LOOPS

In this section, we discuss a peculiar behaviour as experimentally measured in several patterns. Figure S7 presents a representative $|H_C/H_0|$ versus $H_x$, as obtained from the triangle-shaped $AlO_x/Co/Pt$ sample with $\gamma = 22.5°$ and $L=5$ μm, and an exotic behaviour in the different pattern intended with the same shape and materials stacks. As discussed in the main text, due to the different relative orientations of local magnetization between left and right sides, asymmetric slopes with respect to $H_x=0$ are seen in Figure S7a. In Figure S7b, unexpectedly, asymmetric $|H_C/H_0|$ versus $H_x$ of which maximum is shifted from $H_x=0$ is observed. The measured behaviour is beyond the scope of our model, suggesting that further physical consideration should be made to elucidate underlying physics of it.

In the current half-droplet model, it is assumed that the energy barrier for the nucleation is the same regardless of the switching sides. This approximation is reasonable for general cases, however, in some samples e.g., locally damaged by external irritation, or with a local defect where the barrier for the nucleation is highly reduced compared to other sites, huge difference in the energy barrier for the nucleation between the left and right sides can be allowed. In this case, Eq. (S15) is not valid, and it needs to be modified taking into account different reduction parameters for each side:

$$H_{C,L} = \frac{\varepsilon_1 \pi t \left( \sigma_0 \left( \sqrt{1-(H_{in}/H_K)^2} + \left(H_{in}/H_K + \frac{2D}{\sigma}\right)\left(\arcsin(H_{in}/H_K) - \frac{\pi}{2}\right)\right)\right)^2}{4\mu_0 M_s p k_B T \sqrt{1-(H_{in}/H_K)^2}}$$

$$H_{C,R} = \frac{\varepsilon_2 \pi t \left( \sigma_0 \left( \sqrt{1-(H_{in}\cos\varphi/H_K)^2} + \left(H_{in}\cos\varphi/H_K + \frac{2D}{\sigma}\right)\left(\arcsin(H_{in}\cos\varphi/H_K) - \frac{\pi}{2}\right)\right)\right)^2}{4\mu_0 M_s p k_B T \sqrt{1-(H_{in}/H_K)^2}}$$

(S16).

Here, two different reduction parameters $\varepsilon_1$ and $\varepsilon_2$ for the left and right sides, respectively, are used to include the different nucleation energy barriers, which are physically connected to the local reduction in the domain wall energy introduced by reduction in anisotropy and exchange coupling at local defects like dislocation, grain, and roughness. Note that, for the shake of simplicity, we assumed that the reduction parameters are isotropic. Using the different reduction parameters, the coercive field $H_{C,L}$ and $H_{C,R}$ normalized to $H_0$, then, can be rewritten as

$$\begin{cases} H_{C,L}/H_0 = \sigma^2(H_x) / \left[ \sigma^2(0)\sqrt{1-H_x/H_K} \right] \\ H_{C,R}/H_0 = \varepsilon \sigma^2(H_x \sin\gamma) / \left[ \sigma^2(0)\sqrt{1-H_x/H_K} \right] \end{cases} \quad (S16)$$

where the ratio between two reduction parameters $\varepsilon \equiv \varepsilon_1/\varepsilon_2$. Here, $\varepsilon$ is only considered in the coercive field for the right side, assuming that the nucleation process in the absence of the in-plane field is dominated by the left side. By using material parameters of $M_S$=1.49 MJ/m$^3$, $\mu_0 H_K$ =0.45T, $A$=10 pJ/m, $D$=-1.43 J/m$^2$, corresponding to the values extracted in AlO$_x$/Co/Pt (see the main text), and $\varepsilon \simeq 2$, we found the best fit as displayed in Figure S7b, showing quantitatively good agreement with our experimental data. The estimated huge difference in reduction parameter $\varepsilon_1 \simeq 2\varepsilon_2$ between two opposite sides is not fully understood, but we speculate that it may be attributed from non-uniform stacks or local damage from Ar$^+$ irradiation or roughness near side surfaces, implying that samples should be prepared with great care for the measurement.

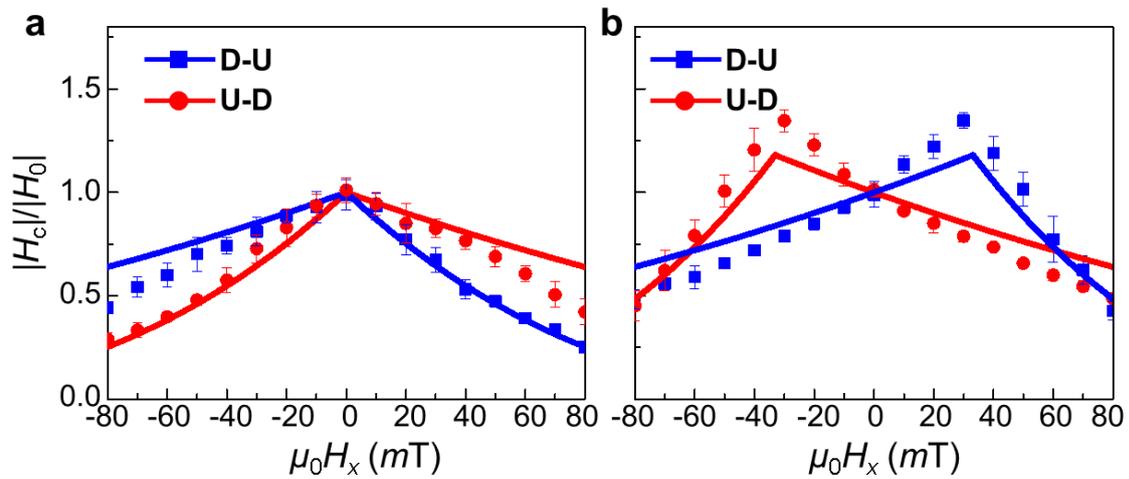

Figure S7. Typical **a** and exotic **b** behaviour of $|H_C/H_0|$ versus $H_x$, as obtained from triangles of $\gamma$= 22.5° and $L$=5μm consisting of AlO$_X$/Co/Pt. The symbols represent experimental data. The solid lines in **a** and **b** are the least-square fit to Eq. (1) in the main text and Eq. (S16), respectively.

## VIII. BLS MEASUREMENT

To validate the reliability of our estimation on $D$, we independently performed BLS measurement. For the measurement, two sample structures of Ta(4)/Pt(4)/Co($t_{Co}$)/Ir(4) and Ta(10)/AlO$_x$(2.5)/Co($t_{Co}$)/Pt(4) (nominal thickness in nm), where the thickness of Co $t_{Co}$ varies from 0-2 nm in a wedge shape, were used in the measurements to observe thickness-dependent $D$. Figure S8a exhibits representative BLS spectra measured in Pt/Co/Ir ($t_{Co}$=1.6 nm) and AlO$_x$/Co/Pt ($t_{Co}$=1.5 nm) under an external in-plane magnetic field $\mu_0 H_{ext}$=0.619 and 0.557 T, respectively. The clear difference in frequency $\Delta f$ between stokes and anti-stokes frequency are observed in both sample stacks, implying existence of DMI. Interestingly, it is seen that the sign of $\Delta f$ is opposite to each other. For Pt/Co/Ir case, the frequency for a Stokes mode is higher than that for an anti-Stokes one, by contrast, for AlO$_x$/Co/Pt, it is lower than that for the anti-Stokes, indicating the opposite sign of $D$ between two sample stacks. This is in consistence with our experimental data from asymmetric hysteresis loops, as discussed in the main text. In Figure S6b, the effective $D$ versus $1/t_{Co}$ as measured from our two sample stacks. To compare the values with those from other stacks, we further plotted $D$ versus $1/t_{Co}$ for three different materials stacks of Ta/Pt/Co/AlO$_x$, Pt/Co/AlO$_x$, and Ta/Ir/Co/AlO$_x$, as reported in our previous papers[6-8], as seen in Figure S8b. From various thickness of Co, we found $D_S$ = 1.61±0.14, and -1.34±0.12 pJ/m in Pt/Co/Ir, and AlO$_x$/Co/Ir, respectively.

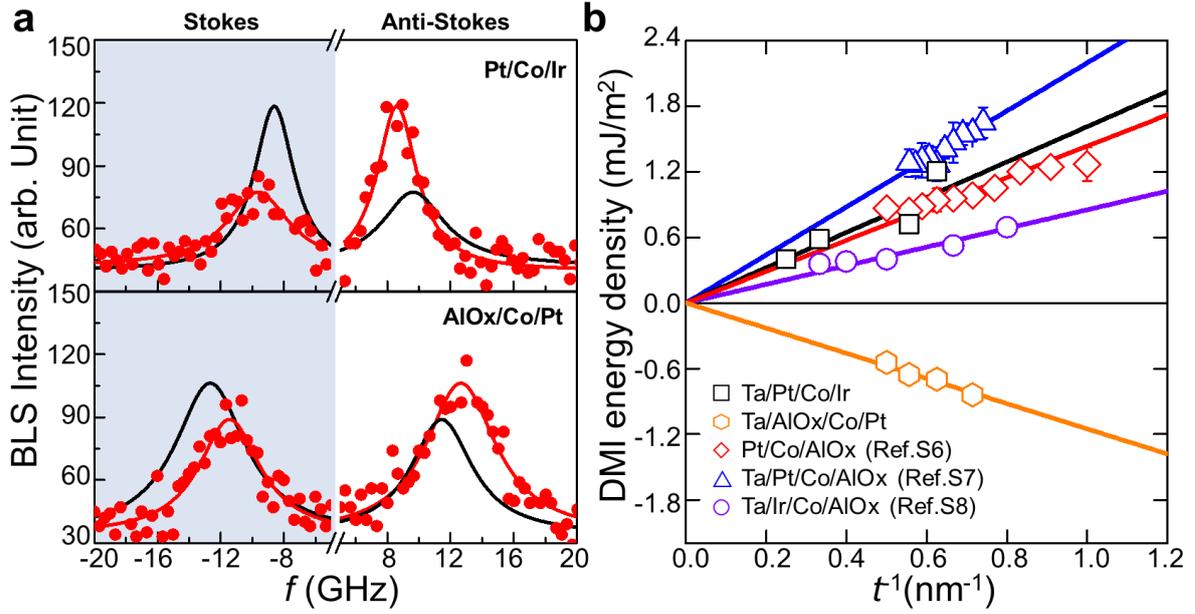

Figure S8. **a** BLS spectra measured in Pt/Co/Ir (top) and AlO$_x$/Co/Pt (bottom) under an external magnetic field $\mu_0 H_{ext}$=0.619 ($t_{Co}$=1.6 nm) and 0.557 T ($t_{Co}$=1.5 nm), respectively, and at an incident angle $\theta$=45° ($k_x$=0.0167 nm$^{-1}$). The red circles and solid lines indicate experimental data and their Lorentzian fits, respectively, The black lines represent mirrored curves of red solid lines with respect to $f$ = 0 GHz. The Stokes and anti-Stokes mode corresponds to peaks in each negative (positive) and positive (negative) frequency region, respectively, for the experimental data. **b** $D$ versus $1/t_{Co}$ for five different materials stacks as measured from BLS.

## IX. SUPPLEMENTARY MOVIES

Movie S1 (S2). Differential Kerr microscopy image of the domain structure as measured in the square-patterned Pt/Co/Ir with $L$=5 μm, under $\mu_0 H_x$=120 mT (-120 mT) and indicated $H_z$ fields. In the reference image, the magnetization is saturated along –z direction. The grey and black colours in the structure represent magnetization pointing down- and up-ward, respectively.

Movie S3 (S4). Differential Kerr microscopy image of the domain structure as measured in the triangle-patterned Pt/Co/Ir with $L$=5 μm, $\gamma$= 30º, under $\mu_0 H_x$=120 mT (-120 mT) and indicated $H_z$ fields. In the reference image, the magnetization is saturated along –z direction. The grey and black colours in the structure represent magnetization pointing down- and up-ward, respectively.

Movie S5 (S6). Differential Kerr microscopy image of the domain structure as measured in the triangle-patterned $AlO_x$/Co/Pt with $L$=5 μm, $\gamma$= 30º, under $\mu_0 H_x$=80 mT (-80 mT) and indicated $H_z$ fields. In the reference image, the magnetization is saturated along –z direction. The grey and black colours in the structure represent magnetization pointing down- and up-ward, respectively.

Movie S7 (S8). Differential Kerr microscopy image of the domain structure as measured in the triangle-patterned $AlO_x$/Co/Pt with $L$=5 μm, $\gamma$= 30º, under $\mu_0 H_x$=80 mT (-80 mT) and indicated $H_z$ fields. In the reference image, the magnetization is saturated along –z direction. The grey and black colours in the structure represent magnetization pointing down- and up-ward, respectively.

# SUPPLEMENTARY REFERENCES